\documentstyle[epsffrag,rotate,graphicx,prl,twocolumn,aps]{revtex}

\newcommand{\bm}[1]{\mbox{\boldmath$#1$}}

\begin{document}
\draft
\twocolumn[\hsize\textwidth\columnwidth\hsize\csname @twocolumnfalse\endcsname

\title{Geometric Laws 
of Vortex Quantum Tunneling}
\author{Uwe R. Fischer$^{1,2}$\cite{email}}
\address{$^1$Institut f\"ur Theoretische Astrophysik, Universit\"at
T\"ubingen, 
Auf der Morgenstelle 10, D-72076 T\"ubingen\\ 
$^2$Low Temperature Laboratory, 
Helsinki University of Technology, 
P.O. Box 2200, FIN-02015 HUT}  

\date{\today}
\maketitle


\begin{abstract}                
In the semiclassical domain the exponent of 
vortex quantum tunneling is dominated 
by a volume which is associated with the 
path the vortex line traces out during its escape from the metastable well.
We explicitly show the 
influence of geometrical quantities  
on this volume by describing point vortex motion 
in the presence of an ellipse. It is argued that for the 
semiclassical description to hold 
the introduction of an additional geometric constraint, the distance
of closest approach, is required. 
This constraint implies that the semiclassical 
description of vortex {\em nucleation} by tunneling 
at a boundary is in general not possible.    
Geometry dependence of the tunneling volume provides
a 
means to verify 
experimental observation of 
vortex quantum tunneling in the superfluid Helium II.  
\end{abstract}

\

\pacs{PACS numbers: 66.35.+a, 67.40.Vs} 

\

] \narrowtext

The question how a singular line with circulating flow around it
enters a fluid in which
friction is absent has pervaded the theory of superfluids for decades.
It is currently believed that 
at the absolute zero of temperature,
a quantized vortex can come into the
superfluid by a 
tunneling process.
In this process, Galilean invariance 
has to be broken because 
there exists no absolute reference frame in the homogenous bulk if the
temperature equals zero.  
The formalism able to describe the quantum mechanical process of tunneling 
depends   
sensitively on the relevant length scales involved.
If the scale after tunneling, 
when  the vortex is able to move freely in the
superfluid, is much larger than the core size of the vortex, we
are able to resort to a hydrodynamic, {\it i.e.} semiclassical
description \cite{volovik}.
If this scale is near the core size $\xi$, a 
description of vortex tunneling requires
knowledge of the dynamic 
many-body structure of the vortex
core, a formidable goal yet to be achieved.
In this paper, we address the question of geometric laws involved in the 
exponent of vortex quantum tunneling in the semiclassical approximation
and investigate the limits imposed on this exponent 
by the shape of the flow obstacle breaking Galilean invariance. 
We show that the usual description of tunneling with a vortex moving in a 
pinning potential generated by some flow obstacle
is limited by a geometric constraint related to the vortex core size. 

The tunneling rate of the vortex 
is proportional to $\exp[-S_e(E)/\hbar]$, where
$S_e(E)\gg \hbar$ is
the Euclidean action as a function of vortex energy $E$\cite{ll}: 
\begin{equation}\label{action}
S_e (E)
= \oint\! d\sigma
\!\oint d{\bm K}\cdot {\bm P}\,.
\end{equation}
Here, ${\bm K}\equiv -i {\bm X(\sigma)}$ is the imaginary collective 
vortex coordinate
and ${\bm P}(\sigma)$ the local vortex momentum per unit $\sigma$-length (with
$\sigma$ the arc length parameter labeling points on the vortex string).
The integral has to be performed along the 
classical vortex path 
with constant energy.
We assume dissipation to be negligible in
the very pure system under consideration. The only driving 
mechanism behind the tunneling transition 
we consider is the superfluid Magnus force at zero temperature.    

We first show in its full generality that the action (\ref{action}) 
in an incompressible fluid is a geometric quantity.
The corrections induced by the 
mass of the vortex 
have been shown to be negligibly small in the
semiclassical limit 
\cite{grisha2,stephen,feigelman}. They will
consequently be neglected here.
  
In analogy to electrodynamics \cite{myself},  
the canonical momentum $\bm P$ in an incompressible superfluid corresponds to  
the vector potential 
and is a gauge dependent vector 
with components
\begin{equation}\label{moment}
P_i= 
\frac{N_v}{N_s}h\rho_0 
\,b_{ij} X'^j\,,
\end{equation}
where $h \rho_0$ is the bulk particle number density 
multiplied by Planck's action quantum $h$, 
$X'^j \equiv \partial X^j/\partial \sigma$ the line tangent. We took 
the quantum of circulation to be $\kappa=(N_v/N_s)h/m$,  with 
$N_v$ the winding number of the vortex and $N_s$ the number 
of real particles in the superfluid boson 
(a Cooper pair has $N_s=2$).
The antisymmetric gauge tensor $b_{\mu\nu}$ of string dynamics \cite{kalb},
a 3+1d 
generalization of the stream function of classical hydrodynamics,
is defined by the dual transformation 
$v^\gamma 
\epsilon_{\gamma\alpha\mu\nu}
=  b_{\mu\nu,\alpha} + b_{\alpha\mu,\nu} +
b_{\nu\alpha,\mu}\,$
from the conservation 
of the four-current $J^\mu=\rho\, v^\mu$ 
(with $\rho = \rho_0$).
Let ${\bm e}_{\it 1},\, 
{\bm e}_{\it 2},\, {\bm e}_\sigma \equiv {\bm X}'$ 
be a righthanded co-ordinate basis on the string.
In  our non-relativistic case, 
we have 
$-\oint\!\!\oint P_a dX^a \wedge d\sigma 
= (N_v/N_s) h\rho_0\int\!\!\int\!\!\int\!
 \sqrt g \, dX^{\it 1}\wedge dX^{\it 2}\wedge d\sigma$\,, 
where $g$ is the determinant of an arbitrary metric in the basis
${\bm e}_{\it a}$, 
${\bm e}_\sigma$ 
($a={\it 1,2}$).
This yields 
for the components of $\bm P $ in the 
directions ${\bm e}_{\it 1}$, ${\bm e}_{\it 2}$ the relation
\begin{equation}\label{p1p2}
\partial_{\it 2} P_{\it 1} -\partial_{\it 1} P_{\it 2}
= \frac{N_v}{N_s} h \rho_0\sqrt g\,.
\end{equation}   
Equation (\ref{p1p2}) represents
the conjugateness of positions and momenta in the 
directions ${\bm e}_{\it 1}$, ${\bm e}_{\it 2}$
in its general form. As a  well-known 
example, if we gauge $P_{\it 2}=0$,
then, in Cartesian co-ordinates, $P_X = (N_v/N_s) h\rho_0 Y$, the
gauge to be used below in the application to the 2d case. 
We are now able to rewrite the action 
(\ref{action}) as
\begin{equation}\label{volume}
\frac{S_e (E)}h =\frac{N_v}{N_s}\, 
\rho_0\,\int\!\!\!\int\!\!\!\int\! \sqrt g\,
d\sigma\, dZ^1 dZ^2 \,.
\end{equation}
The new co-ordinate differentials  
are $dZ^1=\cos\alpha \, dK^{\it 1} + \sin\alpha\, dK^{\it 2}$,
$dZ^2=-\sin\alpha\, dX^{\it 1} + \cos\alpha\, dX^{\it 2}$. The angle
$\alpha (\sigma)$ parameterizes the local (co-ordinate) 
gauge freedom for the momentum of rotations about ${\bm e}_\sigma$, 
contained  in (\ref{p1p2}). 
In what follows, $N_v/N_s\equiv 1$ for simplicity. 
The gauge invariant expression (\ref{volume}) tells us that 
the dimensionless action 
is the number density 
times the volume which is involved
in the tunneling process.
Note that the action is scaled with $h$, not $\hbar=h/2\pi$. 
This {\em Volume Law} is the dominant contribution to the tunneling
exponent in a 
superfluid in which the non-relativistic limit is the physically relevant one. 
The number of particles in the effective
tunneling volume plays the role of the quantum 
number in a Bohr-Sommerfeld type quantization
of (\ref{action}). We stress that this result holds however 
complicated the motion of the vortex string actually 
is as long as we are in the 
semiclassical domain. 

What is of interest to us here are the limits 
which govern the actual value of the tunneling volume
for a given external geometry.     
To tackle the general features of the  
problem, we 
restrict ourselves to the case of point vortices in 2d, where 
we can discard the complicated $\sigma$-dependence in (\ref{volume}).
We are then in the position to use the tools of 
conformal transformation to calculate the energy of the vortex 
as a function of its coordinates 
near some irregular boundary of the superfluid \cite{milne-t}. 
The simplest realistic geometry of interest to us
for this purpose is that of a half-ellipse whose small semi-axis $a$
is in the direction of the flow parallel to the boundary with a much bigger
semi-axis $b\gg a$ perpendicular to it. 
This provides a realistic 
pinning potential directly related to geometrical properties 
of the pinning site in contrast to a 
harmonic shape of the potential with fitting parameters used, {\it e.g.}, in 
\cite{aothouless,stephen}.

We can map 
the readily solvable problem of a vortex in the presence of a half-circle
by the inverse of the Joukowski transformation \cite{milne-t} to that 
of the half-ellipse-vortex geometry in question. The vortex energy 
contains three terms. The 
first term is half the self energy of a vortex pair and
 the second, which gives the pinning potential, stems from the 
two image vortices inside the circle respectively ellipse required to satisfy
the boundary conditions. 
The last one comes from a superimposed external 
flow velocity $u$ at infinity in the negative $x$ direction parallel to the 
boundary.
We use appropriate elliptic co-ordinates defined by 
$x=l\sinh\chi \cos\eta$, $y=l\cosh\chi \sin\eta$, in which choosing
$\chi=\chi_0$ gives us $a=l\,\sinh\chi_0$, $b=l\cosh\chi_0$, with 
$l=\sqrt{b^2-a^2}\approx b \gg \xi$ the overall length scale.
If we normalize the energy as a function of vortex position $\chi_1,\eta_1$   
by $\tilde{E} \equiv {4\pi \, E}/m\rho_0 {\kappa}^2$,
it takes the form
\begin{eqnarray}\vspace*{1em}\label{tildeE}
\tilde{E} = 
\ln\left[\frac{a+b}{\xi}\,\frac{\exp(\chi_1-\chi_0)|\sin\eta_1| \sinh (\chi_1-\chi_0)}{(\sinh^2(\chi_1-\chi_0)+\sin^2\eta_1)^{1/2}}\right]
\nonumber\\
-\frac{4\pi u(a+b)}\kappa\sinh (\chi_1-\chi_0)|\sin\eta_1|\,.
\end{eqnarray}
We cannot solve 
for the path of constant $\tilde E$ analytically 
if we allow for any values of
$\xi,a,b$ and $u$ in the semiclassical domain. 
What we want to show here are 
general features with respect to the geometrical quantities involved 
in the tunneling exponent. 
Hence for the velocity $u$ we consider the limit
$u\ll \kappa/2\pi l \ll v_L$, where $v_L \equiv \kappa/2\pi \xi$ is the 
characteristic velocity associated with the many-body quantum structure 
of the fluid (`Landau' velocity).
The external current is  limited by  $2 u/v_L\ll a/b$, such  
that the velocity
without vortex at the ellipse top remains well below the critical $v_L$.
\begin{center}
\begin{figure}[hbt]
\epsfysize=0.72\textwidth
\epsfxsize=0.43\textwidth
\epsfbox{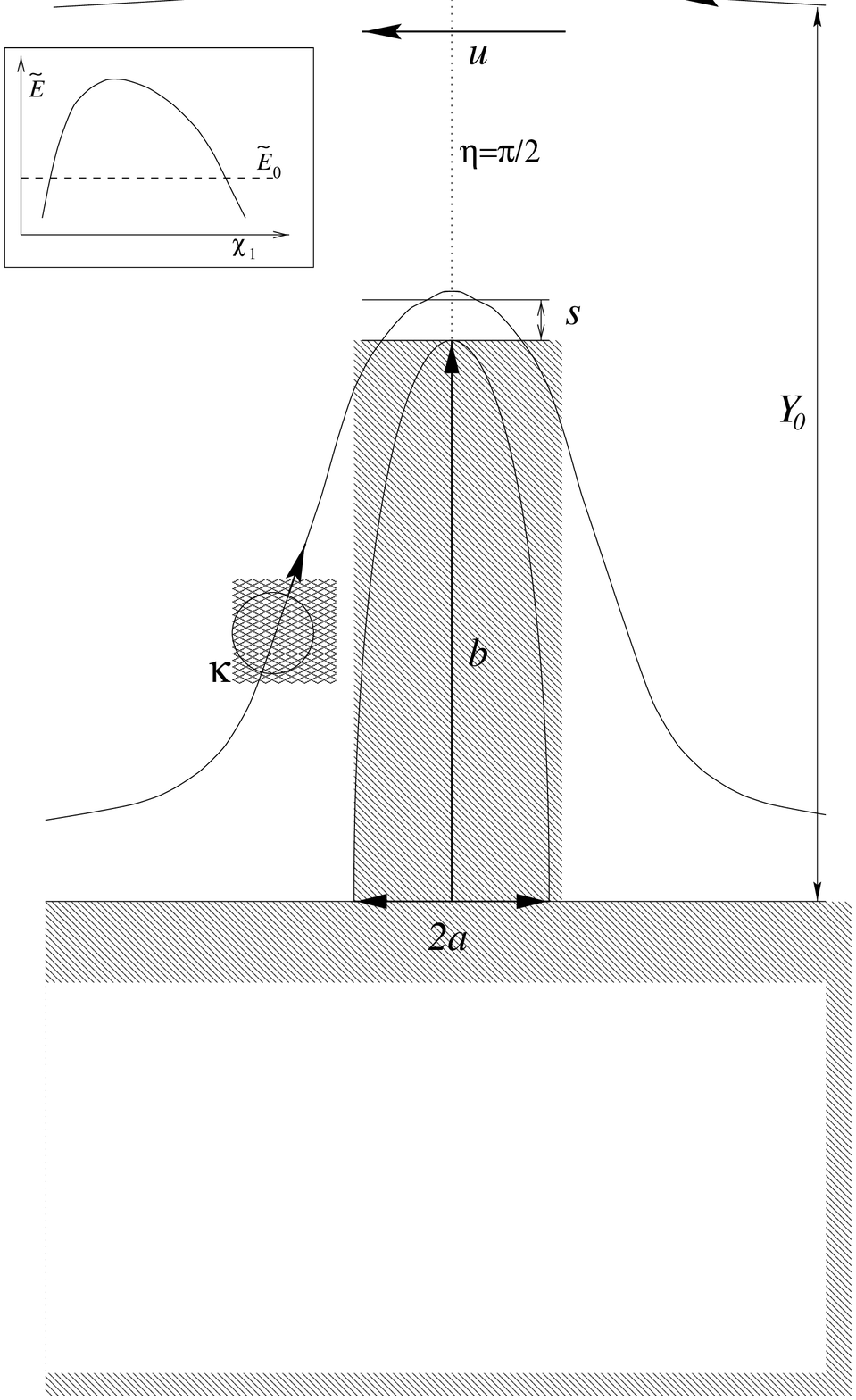}
\vspace*{-10em}
\caption{\label{figure} The two vortex paths of the same constant energy 
$\tilde E_0$.
The vortex on the path passing the ellipse top on a distance slightly 
larger than 
$s\ge\xi$ has to tunnel to the second path, which is at a distance $\simeq Y_0$
from the boundary.  The small inset shows the shape of the barrier from
(\ref{tildeE}) for $\eta=\eta_1=\pi/2$.
}
\end{figure} \vspace*{-1em}
\end{center}
We now come to the crucial point of our argumentation: 
We require that the distance of the vortex to the ellipse top, 
the point of closest approach of the vortex, 
is slightly more than $\xi$, because any smaller distance to the boundary 
would invalidate a description of the vortex in semiclassical terms.
This requirement can be imposed by  
$\delta\chi (\eta_1 =  \pi /2)= s/a$, with $\xi\le s\ll a$,  
and where we set $\delta\chi \equiv \chi_1-\chi_0$. 
The energy $\tilde E\equiv \tilde E_0$ of the paths we consider 
(see Fig. \ref{figure})
is then fixed to be 
\begin{equation}
\tilde E_0\equiv\tilde E (a,b,s,\xi, u) = \ln\left[\frac{a+b}{a}\,\frac s \xi \right]
-\frac{4\pi u (a+b)}{\kappa} \frac{s}{a}\,. 
\end{equation}
For low velocity, 
we can write 
$\exp \tilde E_0 \simeq (b/a) (s/\xi)$, the value to be used in the following.

\begin{figure}[thb]
\includegraphics[width=0.4\textwidth]{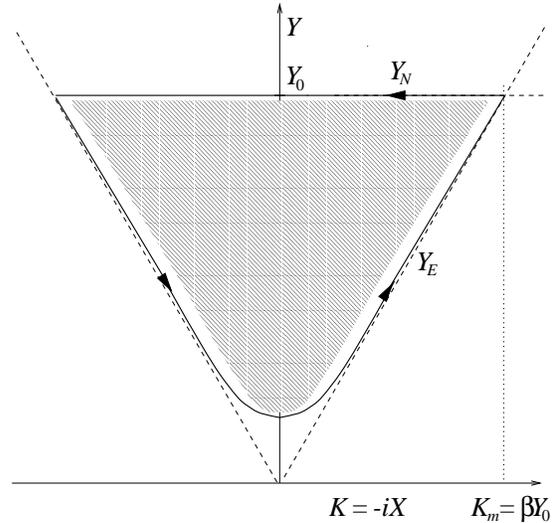}
\caption{\label{area} The area $\simeq \beta Y_0^2=V^{(2)}_N$ 
enclosed by the vortex path in complex space gives the action (\ref{tunnel}) 
in two space dimensions. This area has a lower semiclassical limit  
$(a/b+ \xi/a)Y_0^2$.}  
\end{figure}

There are two vortex paths having the same energy (Fig. \ref{figure}).
One of these 
is that of the vortex 
at the boundary and closely following
the ellipse, for which $\delta\chi\ll 1$. 
We obtain from (\ref{tildeE}), if
$\sin\eta_1\gg\sinh\delta\chi$, $b/a\gg 1$: 
\begin{equation}\label{etanear}
{Y_{\rm E}}
\simeq\frac{\kappa}{4\pi u}\,
\frac1{\delta\chi}\,
\ln\left[
\frac as
\,\delta\chi\right]\,.
\end{equation}
The other vortex path of constant energy $Y_N (\chi_1,\eta_1)$ is 
located far away from the 
ellipse and the boundary.  
The vortex has to tunnel through the barrier represented 
by (\ref{tildeE}) from the ellipse path (\ref{etanear}) to this trajectory. 
In lowest order of the velocity $u$, this path  follows a constant 
distance $Y_0$ to the boundary: 
\begin{equation}\label{etafar}
{Y_{\rm N}}
\simeq Y_0\equiv \frac{\kappa}{4\pi u} 
\ln \left(\frac{\kappa}{2\pi u}\frac {a/b}{s}\right)
\,.
\end{equation}
The two paths (\ref{etafar}) and (\ref{etanear}) do never meet 
provided that $2u/v_L \ll a/b$ holds. 
To describe the tunneling trajectory of constant energy, 
we have to use complex co-ordinates
in the $Z^1,Z^2$ space contained in (\ref{volume}) such that they meet at some
point in this space. We can then obtain a closed escape path. 
To the end of finding the path, 
we return to the original Cartesian co-ordinates and write
the vortex trajectory in the complex ($K=-iX,Y$) space as
\begin{equation}\label{y1K}
Y_{\rm E}^2
=l^2\cosh^2(\chi_0+\delta\chi) + \tanh^{-2}(\chi_0+\delta\chi)K^2\,,  
\end{equation}
where the deviation from the original ellipse $\delta \chi$ is 
in general via (\ref{etanear}) 
itself a function of $K,Y$. In lowest order of $\delta\chi/\chi_0$, 
however, (\ref{etanear}) yields approximately a constant 
$\delta \chi\simeq s/a$ equal to its value at the ellipse top. 
The two trajectories (\ref{etafar}) and (\ref{y1K})
then cross at  
\begin{equation}
|K_m|\simeq 
\left( \frac ab +\frac{s}{a}\right) Y_0\equiv \beta Y_0\,.
\end{equation}
Evaluating the integral (\ref{volume}) leads to the tunneling 
action (see also Fig. \ref{area}):
\begin{eqnarray}
\frac{S_e}{h} & = &  
2\rho_0 \int_0^{K_m}\!\!
(Y_{\rm N}-Y_{\rm E})dK\simeq 
\rho_0 
\beta  Y_0^2
\nonumber\\
 & = & 
\frac 1{16\pi^2}
\rho_0 
\left(\frac ab +\frac{s}a \right)
\left(\frac{\kappa}{u}
\ln\left[\frac{\kappa}{2\pi u}\frac{a/b} {s}
\right]\right)^2\,. \label{tunnel}
\end{eqnarray}
We remind the reader that the result in this simple analytical form 
is valid to lowest order in the small quantities 
$2u/v_L\ll a/b\ll 1$, $s/a\ll a/b$.
But the feature that 
$s$,\,$a$,\,$b$ enter the action 
does also hold for more general values in the parameter space.
We conclude that the semiclassical action 
depends on three geometrical quantities:
The characteristic lengths  
parallel and perpendicular to the flow $a,b$ 
and  the closest approach distance $s$ to the ellipse. 
The quantity $s$ enters because we imposed the condition that the vortex 
must not get nearer to the surface than $\xi \le s$. 
This resulted in an effective
rescaling $\xi\rightarrow \xi\exp \tilde E_0$ 
of the ultraviolet cut-off in the vortex energy
logarithm and a  restriction on the effective sharpness 
$\beta\ge a/b+ \xi/a$ of the ellipse.

As a generalization of (\ref{tunnel}), the semiclassical action 
for a massless vortex in $d$ dimensions will take the form 
\begin{equation} \label{volumeaction}
\frac{S_e}{h}=\rho_0\, \beta\, V^{(d)}_N\,,
\end{equation} 
where $\beta\ll 1$ characterizes the effective dimension of the vortex
escape path, {\it i.e.}\,\,the relative degree to which it is
confined to $n$ dimensions by the presence of an 
asperity which is effectively $n$-dimensional. The quantity $\beta$ is 
bounded from below by the
requirement of semiclassicality for the vortex path at the boundary,
as expounded above for the two-dimensional case.   
The tunneling volume  $V^{(d)}_N$ is that for a vortex escape
path of O$(d-1)$ symmetry \cite{bounce}, which is the highest 
possible symmetry if one preferred direction, namely that of the external 
current, is given. 
In our $d=2$ case of (\ref{tunnel}), we have $n=1$, 
$\beta \simeq a/b +s/a$ and 
$V^{(2)}_N = Y_0^2$.
In three dimensions ($d=3$) 
we have $V^{(3)}_N = (4\pi/3) R_0^3$ for a half-ring 
with radius $R_0$.

The conclusions we have drawn should be able to shed light on the question 
if in experiments 
on the flow of superfluid $^4\!$He through sub-micron orifices 
we are dealing with the quantum tunneling of vortices below about $T=150$ mK 
\cite{jcdavis,ihasAV}.
Geometry dependence provides a tool to verify that the critical velocity 
plateau observed there comes indeed from 
a quantum process and not an alternative process of classical instability, 
{\it e.g.} of the Kadomtsev-Petviashvili type \cite{kuznetsov}, 
is taking place.
In $^4\!$He, the 
velocity $v_L$ is quite large due to the small 
coherence length in the order of the interparticle spacing (it equals 
the Landau velocity $\approx 59$ m/sec at $p=1$ atm 
if $\xi\simeq 2.7$\AA\,\, is taken).   
The observed critical velocities of flow through the orifice on the
temperature independent plateau are of the order of 10 m/sec
\cite{ihasAV,AV3He}. This
leads to materialization radii $R_0$ 
of the order of nanometers 
($R_0$ corresponds to the 2d quantity $Y_0$ in (\ref{etafar})).
A hydrodynamic treatment appears thus justified and there should be a
strong dependence of critical velocities on the surface structure of
the orifices. The simple model of vortex half rings 
situated at plane walls with their momentum axis antiparallel to the flow  
\cite{ihasAV,AV3He}
does not account for the required explicit 
incorporation of Galilean invariance violation.  
The enlargement of the core size 
at boundaries used in this model
to explain observed critical velocities is naturally contained here.
It is represented by the condition that a 
vortex moving in a pinning potential caused by a nontrivial geometry
only exists as a semiclassical vortex able to tunnel through the barrier  
if its distance to the wall 
is everywhere bigger than $\xi$. This results in a 
nonzero tunneling energy $\tilde E_0>0$ of the vortex interpretable as 
an enlargement of $\xi\rightarrow \xi\exp\tilde E_0$. 

The necessity of nonzero tunneling energy is  
responsible for the fact that we do not deal with the {\em nucleation} 
of vortices here, {\it i.e.} their creation from `nothing'. The vortex 
already has to exist 
at the boundary with a small but nonzero 
energy such that it can tunnel to get a free vortex. Hence we 
are dealing with a depinning rather than a nucleation event.    
In the semiclassical limit the description of the nucleation 
of zero energy vortices is in general 
only possible if the curvature radius of the flow obstacle is constant,
that is in the case of $d-1$-dimensional spheres.  
If the curvature radius of the obstacle 
varies, the distance of the zero energy vortex to the boundary varies 
and can get smaller than $\xi$ at the point of closest approach, 
thus violating the requirements of the semiclassical approach. 
 
To conclude, we emphasize again the important fact
that violation of Galilean 
invariance in the direction of the 
flow is a necessary condition for vortex quantum
tunneling to happen at the absolute zero of temperature.   
The tunneling exponent will depend on geometrical properties 
of the vortex path associated with 
curvature scales of the asperities breaking the invariance. That the   
semiclassical description  be valid introduces the     
additional length scale $s$, the closest approach distance to the flow
obstacle.  
The semiclassical theory of vortex quantum tunneling
in a pure  
superfluid is essentially a geometric theory.

I am indebted to Grisha Volovik for stimulating discussions and helpful 
comments.
This work was supported by the Human Capital and Mobility Programme
of the European Union (contract number CHGE-CT94-0069).  
 

\begin{references}
\bibitem{bounce} It has already been pointed out in 
\cite{feigelman} that we do not deal with the O$(d-1)$-bounces of field
theory \cite{kaolee} 
because a massless vortex does not `bounce' back in 
the potential.
\end{references}
\end{document}